\documentclass[letterpaper,twocolumn,10pt]{article}
\PassOptionsToPackage{table}{xcolor}
\usepackage{float}
\usepackage{hyperref}
\usepackage{usenix}
\usepackage{amsmath}
\usepackage{graphicx}
\usepackage{amssymb}
\usepackage{tabularx}
\usepackage{booktabs}
\usepackage{enumitem}
\usepackage{multirow}
\definecolor{MeanBlue}{RGB}{232,243,251}

\begin{document}

\date{}
\title{\Large \bf ActReal: System-Level Mobile Agents Challenge Mobile Automation Detection}
\hypersetup{
  pdftitle={ActReal: System-Level Mobile Agents Challenge Mobile Automation Detection},
  pdfauthor={Mingshuo Wang, Hanqing Guo, Huining Li, Yuliang Fu, Jing Xu, and Chenhan Xu}
}

\author{
{\rm Mingshuo Wang$^{1}$ \quad Hanqing Guo$^{2}$ \quad
Huining Li$^{1}$ \quad Yuliang Fu$^{1}$} \\
{\rm Jing Xu$^{1}$ \quad Chenhan Xu$^{1}$} \\[3pt]
{\rm $^{1}$North Carolina State University \qquad
$^{2}$Indiana University} \\[3pt]
{\small\tt mwang49@ncsu.edu \quad guohan@iu.edu \quad hli83@ncsu.edu} \\
{\small\tt yfu34@ncsu.edu \quad jxu65@ncsu.edu \quad cxu34@ncsu.edu}
}

\maketitle

\begin{abstract}
System-level mobile agents are evolving from fixed scripts into adaptive systems that continuously observe interfaces, reason, and adjust their actions, allowing automated attacks to navigate dynamic UIs and complete complex tasks. Existing applications detect automation using touch trajectories, action timing, and the physical coupling between touch and inertial measurement unit (IMU) signals. However, a privileged system-level agent executor can control both touchscreen input and application-visible sensor delivery, enabling it to jointly generate time-aligned touch and six-axis IMU signals and evade these defenses. We present ActReal, a physical-action attack framework for system-level mobile agents. ActReal converts semantic agent actions into task-valid touch and IMU events using genuine-trajectory adaptation and physics-guided IMU generation. ActReal achieves a mean event-level attack success rate of 77.5\%; even when detectors jointly observe touch and IMU, its attack success rate remains 71.1\%.
\end{abstract}

\section{Introduction}
\label{sec:introduction}

Mobile agents are rapidly emerging as a new paradigm for smartphone interaction. By continuously observing the screen, reasoning about the next action, and replanning in response to changes in the interface, they can adapt to relocated UI elements, transient pop-ups, and different task branches~\cite{zhang2025appagent,wang2024mobile,wen2024autodroid,rawles2025androidworld}. At the same time, increasingly capable mobile agents challenge conventional automation defenses. Unlike fixed scripts, they can interpret interface states, adapt to unexpected changes, and retry failed actions. Recent work shows that agentic vision--language systems can solve 60.7\% of 2,600 challenges across 26 visual CAPTCHA types~\cite{teoh2025captchas}. 
A vendor-commissioned survey further estimated that malicious automation costs enterprises about 4.3\% of online revenue annually~\cite{netacea2024billionbots}.

To continuously detect automated interactions, existing mobile defenses rely on two types of evidence: \textbf{1) touchscreen behavior} and \textbf{2) physical device motion}. \footnote{We exclude CAPTCHA-based defenses because they require users to complete an explicit challenge and therefore affecting the user experience.} 
In touchscreen-behavior-based detection, the intuition is that automated agents interact with the screen differently from human users. Human touch usually contains natural variations in trajectory, speed, contact duration, and timing, while automated execution tends to produce more regular and simplified patterns. For example, Touchalytics showed that these touch dynamics can serve as behavioral biometric signals~\cite{frank2012touchalytics}. More recently, AHB~\cite{zhu2026turing} found that unmodified mobile agents often produce overly linear swipe trajectories, near-zero tap durations, and unusually long pauses caused by model inference.

In physical-signal-based detection, the idea is that real human touch causes small but measurable device motion. When a finger taps, swipes, or presses the screen, the phone typically shows corresponding changes in acceleration and angular velocity. In contrast, conventional automation tools and agents can inject touch events, but they do not create the physical motion that normally accompanies human interaction. BeCAPTCHA~\cite{acien2021becaptcha} and zkSENSE~\cite{querejeta2019zksense} therefore use accelerometer and gyroscope signals, together with touch behavior, to distinguish software-generated interactions from real human input. However, these methods rely on a critical assumption: \emph{the IMU signals are real and cannot be faked by the agent.}

As agents become increasingly integrated into mobile operating systems and smartphones, this assumption is facing substantial challenge. 
For example, Android Computer Control and Gemini have demonstrated system-integrated interface perception and application operation~\cite{androidComputerControl,googleGeminiMultistep}, while a recent analysis of the Doubao Mobile Assistant showed that a system-level agent can use privileged permissions, virtual displays, and hidden system APIs to observe target applications and inject input~\cite{zou2026blind}. These privileged capabilities of agent raise a new risk: a system-level executor may control both touch input and application-visible sensor signals, allowing agent actions to appear human.

To further investigate this security risk, we develop ActReal, an attack framework that enables system-level mobile agents to mimic human physical interaction. ActReal operates between the agent planner and the Android executor. It converts the agent’s semantic actions such as target locations, movement directions, or requested text into time aligned touch trajectories and six-axis IMU signals, and delivers both to the target application through a system-level executor. Consequently, the agent can mimic both touch and device motion. This allows it to bypass both touchscreen-behavior-based detection and physical-signal-based detection.
 
\begin{figure}[t]
  \centering
  \includegraphics[width=0.95\columnwidth,trim=234 42 326 42,clip]{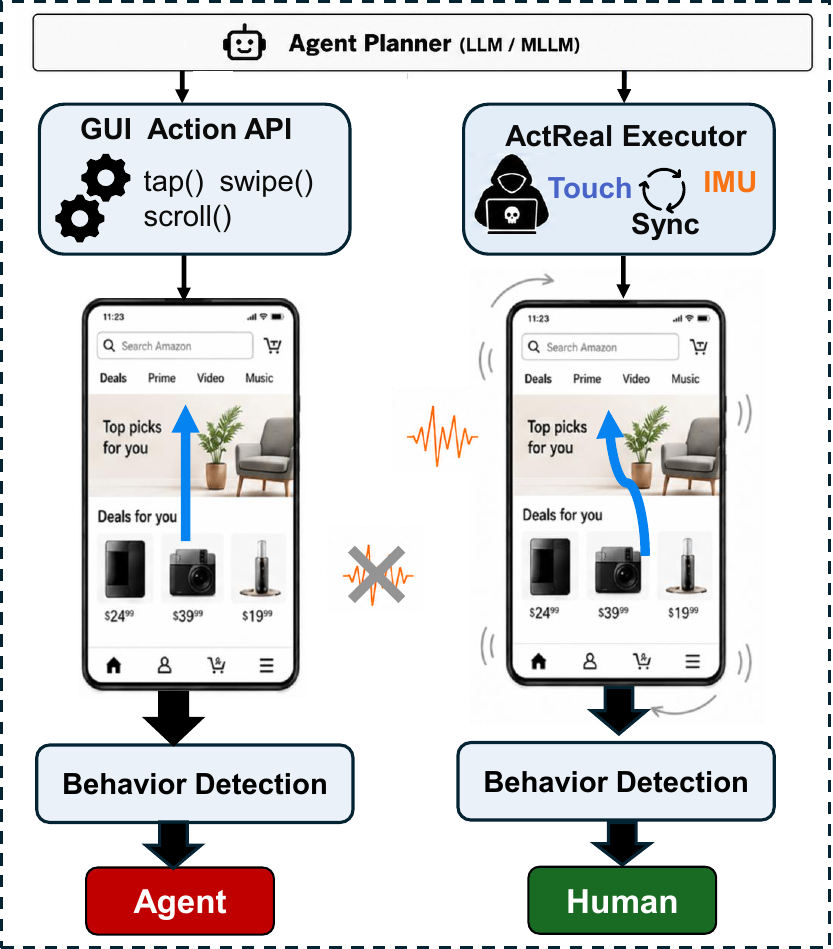}
  \caption{Overview of ActReal's core idea. Conventional GUI automation uses Android API, whereas ActReal synchronizes
  touch and IMU signals so that behavioral detectors may classify the
  interaction as human.}
  \label{fig:actreal-overview}
\end{figure}

Figure~\ref{fig:actreal-overview} contrasts this joint realization with conventional GUI-action injection.

Implementing ActReal presents three main challenges. 

\noindent\textbf{1. Lack of human-like touch execution.}
To bypass touch-behavior-based detection, an agent needs to perform each action with human-like touch patterns. However, existing mobile-agent executors are mainly designed to complete the requested action. They can execute commands such as Tap and Swipe, but they do not provide realistic low-level touch behavior, such as natural trajectories, contact duration, movement timing, and multi-finger gestures. For example, an automated swipe may follow a simple straight path, while some agent interfaces do not support a two-finger pinch at all~\cite{wang2025mobile,zhang2025appagent,li2026mobileuse}. \textit{ActReal therefore needs to generate low-level touch events that preserve the intended actions while reproducing human interaction patterns.}

\noindent\textbf{2. Lack of realistic device-motion generation.}
To bypass physical-signal-based detection, the agent needs to produce IMU signals that look like real device motion during human operations and match the corresponding touch actions. However, generating signals with a similar overall distribution is not enough. Existing detectors can examine temporal changes, frequency patterns, cross-axis relationships, and the alignment between touch and IMU. Generic time-series generation models are not designed to preserve all of these properties at the same time~\cite{yuan2024diffusion,naiman2024utilizing}. \textit{ActReal therefore needs to generate realistic six-axis IMU signals that remain synchronized with the humanized touch events.}

\noindent\textbf{3. Lack of efficient online humanization.}
To bypass continuous behavioral detection, the agent needs to generate human-like touch and IMU signals without adding noticeable delay. However, mobile agents already have long and variable pauses caused by screen perception and model reasoning~\cite{zhu2026turing}. Generating realistic signals during each action can introduce additional latency and make these pauses even more abnormal. \textit{ActReal therefore needs to reduce online generation overhead while also making the long gaps between agent actions appear more like human interaction.}

To address these challenges, ActReal employs three mechanisms. First, ActReal uses a unified system-level execution wrapper to realize agent intended actions as low-level touch and IMU events. To support heterogeneous action spaces across different agents, the wrapper first normalizes agent outputs into five basic physical action classes: tap, scroll, swipe, pinch, and keystroke. Then, the wrapper generates and delivers the corresponding event sequences, overcoming the limited action expressiveness of higher-level interfaces. 
Second, ActReal humanizes both touch behavior and device motion. For touch, ActReal starts from genuine human trajectories and adapts them to the current target location and action geometry, while preserving natural variations in movement and timing. For IMU, ActReal uses diffusion models to generate six-axis motion signals and applies physics-guided constraints to reduce unrealistic temporal, frequency, and cross-axis patterns. ActReal places the touch and IMU signals in the same ActionBundle and aligns the clocks of their two Android event paths before execution. Both signals are then mapped onto the same action timeline, giving them the same start time and duration. 
Finally, ActReal reduces online overhead by generating and caching IMU signals in advance, while adapting touch trajectories at execution time. In addition, when long pauses occur because the agent is perceiving the screen or reasoning about its next action, ActReal provides background IMU signals and inserts randomized filler taps to make these pauses less distinguishable from normal human interaction. Our contributions are summarized as follows:
\begin{enumerate}
  \item \textbf{A security challenge to mobile automation detection.} We
  identify a key limitation of existing mobile automation defenses: when a system-level agent controls both Android's input-dispatch path and application-visible sensor-delivery path, touch--IMU consistency is no longer sufficient to establish human origin. We formalize this threat under a system-level mobile-agent attack model.
    \item \textbf{Physics-guided human-like interaction generation.} We design a
physics-guided method for generating human-like touch and IMU signals. For
touch, ActReal adapts genuine trajectories to the requested action geometry
while preserving their human motion characteristics. For IMU, ActReal performs
five-shot target-user-conditioned generation and further uses complementary critics
and conflict-aware training to reduce detectable artifacts while preserving
realistic signal structure.

    \item \textbf{A system-level touch--IMU execution pipeline.} We design an
execution pipeline that translates outputs from different mobile agents into
low-level actions and coordinates the delivery of touch and IMU events. The
pipeline supports multiple action types, reduces online overhead through IMU
pre-generation and caching, and maintains continuous sensor activity during
long agent-reasoning gaps using background IMU signals and filler taps.

    \item \textbf{A comprehensive security evaluation.} We evaluate ActReal across five action classes, three observation modalities, and six attack-aware detectors. ActReal achieves a mean event-level ASR of 77.5\% overall and 71.1\% under joint touch--IMU observation. We further validate ActReal in real-device settings using three mobile-agent frameworks and two Android phone models, where the generated events achieve ASRs ranging from 48.1\% to 95.3\% across different action--modality pairs.
\end{enumerate}

\section{System and Threat Model}
\label{sec:threat-model}

\subsection{Attack Scenario and Trust Boundary}
\label{sec:event-delivery}

\noindent\textbf{Target system.}
ActReal targets mobile applications that use passive behavioral detection to identify automated interactions. These applications continuously observe touchscreen behavior, interaction timing, and device motion during normal use. This type of detection is suitable for content, social, shopping, and gaming applications because it does not require users to complete an explicit challenge. ActReal does not target high-assurance applications, such as banking or payment services, where touch and IMU are only part of a larger security system.

\noindent\textbf{Attack scenario.}
We consider a system-level mobile agent that performs tasks on behalf of a user. The agent has a privileged executor that can inject touchscreen input and control the IMU signals delivered to the target application. ActReal uses these capabilities to replace normal automated execution with human-like touch events and matching device-motion signals.
If the attack succeeds, the target application observes touch and IMU signals that appear to come from a human user. As a result, the agent can bypass the application's passive behavioral detector and continue automated operations without being identified as automation.

\subsection{Attacker Capabilities}
\label{sec:attacker-capabilities}
We assume that the attacker controls a mobile-agent executor with system-level
privileges that can control input-event dispatch and the
motion-sensor values delivered to the target application. Specifically, the
executor can observe the screen, inject touch events, and replace the
accelerometer and gyroscope sample values visible to the application. This
assumption is supported by two complementary platform capabilities. The
AutoAction component of Doubao Mobile Assistant can use
\texttt{INJECT\_EVENTS} to inject screen input events~\cite{zou2026blind}.
Meanwhile, the Android VirtualDevice framework shows that a privileged system
component can provide applications with controlled virtual accelerometer and
gyroscope data through the standard \texttt{SensorManager} interface.

As described in Section~\ref{sec:method-injection}, the ActReal prototype uses
a persistent multi-touch \texttt{uinput} device to implement touch injection
and intercepts \texttt{SensorEventQueue::read} within the target application
process via a native hook to replace IMU values. Under the threat model
considered in this paper, the attacker is permitted only to control input
dispatch and sensor sample values and is not permitted to modify the target APK
file, detector logic, features, model parameters, thresholds, detection scores,
final decisions, or server-side state.

\subsection{Defender Capabilities}
\label{sec:defender-capabilities}

We consider two types of behavioral defenders: a passive behavioral defender and an attack-aware adaptive defender. For the passive behavioral defender, the target application continuously collects signals during normal user interaction. It obtains touch trajectories, action duration, and inter-action intervals from \texttt{MotionEvent}, and collects accelerometer and gyroscope signals through Android \texttt{SensorManager}. The defender can analyze touchscreen behavior, device motion, and the consistency between touch and IMU signals.

For the attack-aware adaptive defender, we consider a stronger setting in which the defender knows how ActReal works and includes ActReal-generated samples when training the detector. The detection model and threshold are selected using independent validation data and fixed before testing. However, the defender does not know the genuine reference trajectories used by ActReal for unseen users. In both settings, we consider only touch and IMU signals that applications can collect passively. We exclude microphone and camera signals because they require additional permissions or explicit user consent.

\subsection{Attack Goals}

ActReal aims to satisfy three conditions while preserving the task semantics of
the agent. First, \textbf{task validity}: the generated taps, swipes, pinches,
or keystrokes must satisfy the current task constraints, and ActReal must not introduce unintended application-state changes or
alter the intended task outcome. Second, \textbf{execution continuity}: signal
generation and delivery should not introduce noticeable additional delays and
should maintain continuous behavioral signals during the intervals caused by
agent perception and reasoning. Third, \textbf{behavioral evasion}: ActReal
should achieve a high attack success rate for generated touch signals, six-axis
IMU signals, and their joint observations; the target application should
classify the generated interactions as human in a high proportion of attack
attempts.

\section{ActReal Design}
\label{sec:method}
\begin{figure*}[t]
  \centering
  \includegraphics[width=\textwidth]{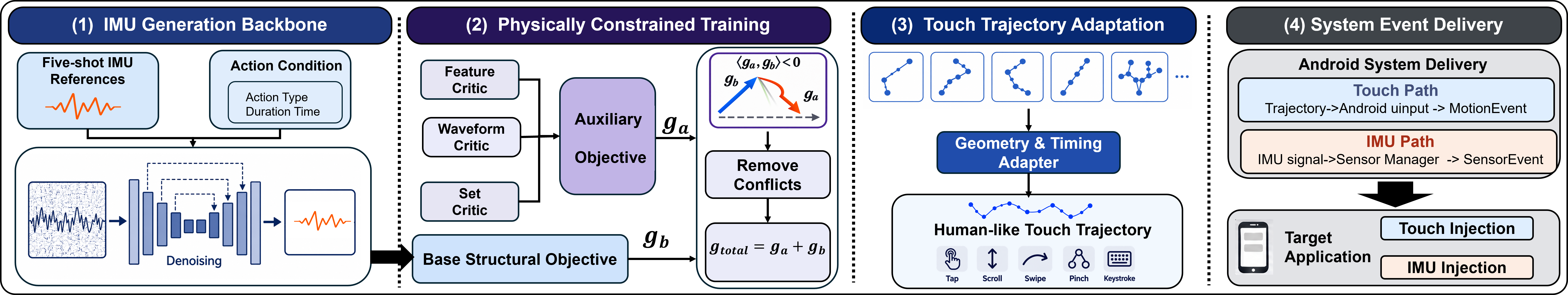}
  \caption{Overview of ActReal's design. The IMU generation backbone is
  conditioned on five target-user reference samples and action metadata; physically
  constrained training combines complementary critics with conflict-aware
  updates; the resulting signals are paired with adapted touch trajectories
  for system-level delivery.}
  \label{fig:method-overview}
\end{figure*}

ActReal transforms a mobile agent's semantic action into coordinated human-like touch and IMU signals. Given an action type and its task constraints, ActReal first maps the action into one of five physical interaction classes. It then adapts genuine touch references to the current task geometry and generates the corresponding six-axis IMU signals. Finally, ActReal aligns the two streams in time and delivers them through the Android execution layer. Figure~\ref{fig:method-overview} summarizes these components and their data flow.

ActReal consists of four main components. It first translates an agent’s high-level actions into a unified set of mobile interactions (\S\ref{sec:method-actions}). Given these actions, ActReal generates user-conditioned IMU signals (\S\ref{sec:method-imu}) and further improves their realism through critic-based and structure-preserving objectives with conflict-aware optimization (\S\ref{sec:method-adv}). It then adapts human touch trajectories to the target action geometry (\S\ref{sec:method-trajectory}) and delivers the resulting touch and IMU signals through Android’s event paths (\S\ref{sec:method-injection}).

\subsection{Action Model}
\label{sec:method-actions}
Existing mobile agents expose different action abstractions, which commonly include tapping, swiping, scrolling, and typing~\cite{zhang2025appagent,wang2024mobile,rawles2025androidworld}.
 These action spaces cover point selection, single-pointer movement, and text entry, but generally omit multi-pointer input.  When an interface requires pinch-to-zoom, rotation, or another coordinated gesture, an agent may choose the correct operation yet
lack a primitive with which to express it.  For \texttt{type(text)}, the agent supplies only the requested string and produces neither the corresponding physical keystroke nor device motion, making agentic inputs easily distinguishable from human operations. These limitations arise
from a common interface abstraction: the agent specifies \emph{what} to do and
leaves \emph{how} the action occurs to the executor. Existing simplified
executors cannot fully realize tasks that require multi-pointer input or physical per-key entry. The target application, meanwhile, observes touch coordinates, timestamps, and pointer state delivered by Android
\cite{androidMotionEvent}, together with device motion during the interaction.

We therefore design ActReal's action model to bridge the gap between an agent's semantic actions and the physical behavior observable to the application. ActReal's action taxonomy is primarily derived from the interaction types and synchronized touch--IMU data provided by HMOG~\cite{sitova2015hmog}. Specifically, we map and extend agent operations into five classes: tap, scroll, swipe, pinch, and keystroke. These five classes cover the major interaction forms required by common mobile GUI agent tasks, including selecting UI elements, scrolling through content, performing directional gestures, manipulating multi-touch interfaces, and entering text. Among them, pinch represents multi-pointer interaction, while keystroke represents a complete typing episode. For keystrokes, the deployment environment additionally provides a character-to-coordinate map for the current keyboard. ActReal first expands the requested text into a sequence of key locations, then uses fixed few-shot references to determine the key-hold duration, inter-key pause, and within-key motion of each key, while generating the corresponding IMU signal for the same keystroke. Table~\ref{tab:action-model} summarizes the semantic inputs for each action class and the physical behavior completed by ActReal.

\begin{table}[t]
  \centering
  \small
  \caption{ActReal's semantic-to-physical action mapping.}
  \label{tab:action-model}
  \begin{tabularx}{\columnwidth}{@{}lX@{}}
    \hline
    Action & Agent input $\rightarrow$ ActReal output \\
    \hline
    Tap & Target $\rightarrow$ landing point, dwell, local motion \\
    Scroll & Direction/distance $\rightarrow$ drag path, endpoint, duration \\
    Swipe & Direction/endpoints $\rightarrow$ stroke, duration, release velocity \\
    Pinch & Zoom/region $\rightarrow$ centroid trajectory, pointer-count state \\
    Keystroke & \texttt{type(text)} $\rightarrow$ key locations, key-hold durations/inter-key pauses, paired IMU \\
    \hline
  \end{tabularx}
\end{table}

\subsection{IMU Generation Backbone}
\label{sec:method-imu}

\paragraph{Fixed-length input representation.}
Action durations vary widely in real smartphone interactions: a tap may last
only tens of milliseconds, whereas a scroll or swipe may last for more than a
second. Generative models, however, are typically trained with fixed-size inputs
and outputs. Rescaling every event to a common length changes the frequency and
phase of its IMU signal, while treating padded regions as part of the action
introduces artificial quiescent intervals. ActReal therefore trains separate
action-specific models for tap, scroll, swipe, and pinch, and represents
variable-duration actions within a fixed carrier window for each action.

ActReal constructs its IMU generation corpus from the paired action records,
accelerometer traces, and gyroscope traces in HMOG~\cite{sitova2015hmog}, with
all IMU traces resampled to a sampling rate of $f_s=100$~Hz. For each action
type, we determine the carrier capacity from its HMOG duration distribution.
After removing duration outliers, we take the 95th percentile of the remaining
events and add a 100-ms allowance before and after the action. This yields
carrier lengths of 350~ms for tap, 1.79~s for scroll, 1.67~s for swipe, and
1.16~s for pinch. Given a requested duration $d$, ActReal sets the active interval to
$n=\operatorname{round}(f_s d)$ samples and marks these samples with an action
mask. The model always generates a fixed-length six-axis IMU window, while the
masked active interval expands or contracts with the requested duration. For
example, the scroll model represents a 400-ms action with 40 active samples and
a 1.2-s action with 120 active samples. Requests exceeding the carrier capacity
are truncated and marked as partial events. However, a keystroke contains multiple contacts rather than a single continuous gesture. ActReal generates a 100-Hz six-axis IMU sequence conditioned on the complete keystroke, as described below.

\paragraph{Five-shot target-user conditioning.}
ActReal trains structurally identical conditional diffusion models for different action types and uses five six-axis IMU samples from the target user as \textbf{five-shot user references}. The model first encodes the five reference samples individually and aggregates them into a user-level representation that captures the target user's characteristic motion patterns. This user representation is then combined with action conditions, such as the requested action duration and device orientation, to condition the denoising network and guide the generation of six-axis IMU sequences that match both the target user and the requested action. All model parameters are pretrained on the HMOG training users; therefore, adapting to a new target user requires only five reference samples and does not involve model retraining.

We adopt DDPM~\cite{ho2020denoising} as the underlying generative framework. During training, DDPM progressively adds Gaussian noise to real IMU sequences and learns to reverse this process conditioned on the user condition derived from the five target-user reference samples together with the action conditions. During generation, the model starts from random noise and progressively denoises it under the same conditions to produce a complete six-axis IMU sequence. To reduce offline generation cost, ActReal uses DDIM for sampling, which reuses the same trained model while requiring fewer denoising steps.

For keystrokes, concatenating independently generated fixed-length windows can introduce discontinuities at window boundaries, which may be exposed by detectors through adjacent-sample jump features. Therefore, ActReal constructs a lightweight signal adapter from the target user's five keystroke references. Given a requested keystroke, the adapter synthesizes the complete six-axis IMU sequence in one pass by combining the user's baseline motion, keystroke impacts, and residual motion. The touch and IMU generators realize the same keystroke, ensuring temporal alignment across the two modalities while avoiding discontinuities caused by concatenating fixed-length windows.

\paragraph{Offline IMU Caching and Online Execution.}
ActReal generates IMU signals offline before task execution and organizes them into a local cache according to action type and duration. For each action type, we enumerate the valid durations observed in the cleaned HMOG training data at a sampling granularity of 100~Hz and generate two IMU samples for each duration. Each cache ultimately contains 34 tap, 326 scroll, 302 swipe, and 212 pinch samples, for a total of 874 IMU sequences, occupying approximately 7.09~MiB on average. In this way, the computationally expensive diffusion sampling is performed only during offline cache construction and does not lie on the agent's online execution path.

During actual execution, the agent framework first generates the next action request based on the current interface state and task, including information such as the action type, target location, and action duration. ActReal then receives this request and selects a matching IMU sequence from the local cache based on the action type and duration, while adapting the corresponding touch trajectory according to the target location and action geometry specified by the agent. For example, for a scroll request with a specified duration and start/end positions, ActReal selects an IMU sequence of the corresponding duration from the scroll cache and maps the pre-generated touch trajectory to the requested start and end positions.

\subsection{Physically Constrained Training}
\label{sec:method-adv}
The base conditional diffusion model can generate IMU signals from the requested action duration and five target-user reference samples, but the standard noise-reconstruction objective does not directly constrain the statistical, temporal, and cross-axis patterns used by behavioral detectors. ActReal therefore introduces multiple complementary critics to reduce these detectable artifacts. We further find that directly combining adversarial and diffusion objectives can create training conflicts, where critic-driven updates improve realism but disrupt the signal structure learned by the diffusion model. To address this, ActReal adds structural constraints and conflict-aware gradient updates to preserve signal quality while benefiting from adversarial training.

\paragraph{Multi-view adversarial critics.}
Let $x\in\mathbb{R}^{T\times6}$ denote a six-axis IMU window after per-channel
normalization over the training users, $m^v$ its validity mask, and $c$ the
action condition.  The condition encodes device orientation, the fraction of
the carrier occupied by the action, and the valid-window fraction.  All three
critics receive the same condition but judge the signal through statistical
features, the normalized waveform, and its relationship to the five target-user
reference samples, respectively.

\emph{Feature critic.}
The feature critic maps the six-axis sequence to behavioral statistics through
a differentiable extractor $\phi(x,m^v)$. We use two feature representations: a basic representation and a richer representation. The basic representation includes
per-channel mean, standard deviation, RMS, adjacent-sample differences, and
frequency-band energy. The rich representation used for scroll, swipe, and
pinch additionally includes quantiles, skewness, kurtosis, a power-weighted
mean-frequency statistic, and within-accelerometer and within-gyroscope axis
correlations.  Concatenating
the features with the action condition, a two-hidden-layer MLP with spectral
normalization produces the realism score in \eqref{eq:feature-critic}:

\begin{equation}
  \label{eq:feature-critic}
  D_{\mathrm{feat}}(x,c)
  = \operatorname{MLP}_{\mathrm{feat}}
    \bigl([\phi(x,m^v),c]\bigr).
\end{equation}

Tap uses a compact 73-dimensional representation; scroll, swipe, and pinch use
the 214-dimensional rich representation.  This critic constrains the generator
in the time-, frequency-, and channel-statistic views commonly available to a
behavioral detector.

\emph{Waveform critic.}
Statistical features may miss local temporal patterns, so the waveform critic directly processes the normalized six-axis IMU sequence. ActReal concatenates the waveform with the validity mask and the time-expanded action condition:

\begin{equation}
  \label{eq:waveform-input}
  h_0 = [x,\;m^v,\;\operatorname{repeat}(c,T)].
\end{equation}

Here, $x\in\mathbb{R}^{T\times 6}$ is the normalized six-axis IMU sequence,
$m^v$ indicates which time steps contain valid sensor data, and
$\operatorname{repeat}(c,T)$ replicates the action condition $c$ across all
$T$ time steps so that the temporal network can access the same condition at
every position.

A stack of one-dimensional convolution layers then extracts local and
longer-range temporal patterns from the sequence. The resulting representation
is pooled over valid positions and mapped to a scalar realism score:

\begin{equation}
  \label{eq:waveform-critic}
  D_{\mathrm{wave}}(x,c)
  = \operatorname{Linear}\!\left(
      \operatorname{MaskedPool}\!\left(
        \operatorname{Conv1D}(h_0),m^v
      \right)
    \right).
\end{equation}

By operating directly on the waveform, this critic captures temporal artifacts
such as abrupt adjacent-sample changes and local oscillations that may be missed
by summary statistics.

\emph{Set critic.}
The feature and waveform critics evaluate whether a generated event resembles
human motion in general, but they do not explicitly enforce consistency with
the target user. The set critic therefore compares each candidate with the
target user's five reference samples. Let
$R=\{r_1,\ldots,r_5\}$ denote the reference set. ActReal summarizes these
references by their feature mean and variation:

\begin{equation}
  \label{eq:reference-statistics}
  \mu_R = \frac{1}{5}\sum_{i=1}^{5}\phi(r_i),
  \qquad
  \sigma_R =
  \sqrt{\frac{1}{5}\sum_{i=1}^{5}
  \left(\phi(r_i)-\mu_R\right)^2 }.
\end{equation}

For a candidate $x$, we extract $z=\phi(x)$ and construct

\begin{equation}
  \label{eq:set-representation}
  q_R(x)
  =
  [\mu_R,\;z,\;|\mu_R-z|,\;
   \mu_R\odot z,\;\sigma_R,\;c].
\end{equation}

Here, $\mu_R$ represents the target user's typical feature pattern,
$\sigma_R$ describes the variation among the five references, and $z$
represents the candidate. The difference and element-wise product capture
how the candidate relates to the target-user reference distribution.
The set critic then maps this representation to a realism score:

\begin{equation}
  \label{eq:set-critic}
  D_{\mathrm{set}}(x,R,c)
  =
  \operatorname{MLP}_{\mathrm{set}}\!\left(q_R(x)\right).
\end{equation}

During training, genuine events from the target user are treated as positive
examples, while generated events are treated as negative examples. We also
use genuine events from other users as mismatched negatives, encouraging the
critic to distinguish target-user consistency.

\paragraph{Structural constraints and protected updates.}
Although the adversarial objective helps reduce detectable artifacts, its updates may disrupt the signal structure already learned by the diffusion
model. ActReal therefore introduces structural constraints and limits the influence of adversarial updates at the gradient level.

Let $\hat{x}_0$ denote the clean waveform predicted by the diffusion model,
$x$ the corresponding genuine waveform, and $m^v$ the validity mask indicating
valid sensor positions. We define the base objective as

\begin{equation}
  \label{eq:base-objective}
  \mathcal{L}_{\mathrm{base}}
  =
  \mathcal{L}_{\epsilon}
  + \lambda_{\Delta}\mathcal{L}_{\Delta}
  + \lambda_F\mathcal{L}_F,
\end{equation}

where $\mathcal{L}_{\epsilon}$ is the standard diffusion noise-prediction loss.
The second term preserves local temporal changes by matching adjacent-sample
differences between generated and genuine signals:

\begin{equation}
  \label{eq:adjacent-consistency}
  \mathcal{L}_{\Delta}
  =
  \operatorname{mean}_{t}
  \left[
    m_t^v m_{t-1}^v
    \left\|
      (\hat{x}_{0,t}-\hat{x}_{0,t-1})
      -(x_t-x_{t-1})
    \right\|_2^2
  \right].
\end{equation}

This constraint discourages abrupt jumps and excessive smoothing in the
generated waveform. We further preserve frequency-domain structure using

\begin{equation}
  \label{eq:spectral-consistency}
  \begin{aligned}
    \mathcal{L}_F
    &= \operatorname{mean}
       \left|S(\hat{x}_0)-S(x)\right|, \\
    S(y)
    &= \log\!\left(
       1+\left|\operatorname{rFFT}(m^v\odot y)\right|
       \right).
  \end{aligned}
\end{equation}

This term encourages the generated and genuine signals to have similar
frequency characteristics. Together, these objectives preserve both local
temporal continuity and the overall spectral structure of human motion.

In addition to the three critics, ActReal uses a feature-matching term
$\mathcal{L}_{\mathrm{fm}}$ to align differentiable statistical features
between genuine and generated IMU batches. We combine this term with the adversarial loss to form the auxiliary objective,

\begin{equation}
  \label{eq:aux-objective}
  \mathcal{L}_{\mathrm{aux}}
  =
  \lambda_{\mathrm{adv}}\mathcal{L}_{\mathrm{adv}}
  +
  \lambda_{\mathrm{fm}}\mathcal{L}_{\mathrm{fm}}.
\end{equation}

The generator jointly optimizes the base structural objective
$\mathcal{L}_{\mathrm{base}}$, which combines the diffusion loss with temporal
and spectral constraints, and the auxiliary objective
$\mathcal{L}_{\mathrm{aux}}$, which uses adversarial and feature-matching losses
to reduce detectable differences between genuine and generated signals. Let
$g_b=\nabla_\theta\mathcal{L}_{\mathrm{base}}$ and
$g_a=\nabla_\theta\mathcal{L}_{\mathrm{aux}}$. Inspired by projected conflicting
gradients (PCGrad)~\cite{yu2020gradient}, when
$\langle g_a,g_b\rangle<0$, ActReal removes from $g_a$ only the component that
opposes $g_b$. It then rescales the conflict-resolved auxiliary gradient if its
norm exceeds a fixed fraction of $\lVert g_b\rVert_2$. Denoting the auxiliary
gradient after conflict removal and norm control by $\bar{g}_a$, the final
generator update is

\begin{equation}
  \label{eq:generator-update}
  g_{\mathrm{total}} = g_b + \bar{g}_a.
\end{equation}

This update retains non-conflicting improvements in realism while preventing
auxiliary optimization from overwhelming the base structural objective.

\subsection{Touch Trajectory Adaptation}
\label{sec:method-trajectory}
Unlike IMU, which can be pre-generated for a finite set of action types and durations, touch trajectories depend on runtime geometry, including the target
location, start and end points, and control boundaries. This continuous geometric space makes it impractical to cache a trajectory for every possible
configuration, while online diffusion sampling would introduce additional latency. Furthermore simple replay is also insufficient because it preserves the original
coordinates, whereas linear interpolation can distort the local variations of
human motion. ActReal therefore stores five genuine touch references for each target user and action type and adapts one of them to the requested geometry at
runtime. After transformation, ActReal verifies that the trajectory reaches the requested target, preserves the required temporal order, and remains within
the screen bounds.
\paragraph{Tap.}
ActReal fixes DOWN inside the target control and carries the reference's small
lift drift to UP while retaining relative timing.  If the reference motion would
exceed the screen or the touch-slop threshold, ActReal uniformly shrinks the complete excursion.

\paragraph{Scroll and swipe.}
Both actions use the same geometric transform.  Let $p_i$ be the $i$th point in
the reference trajectory and $u_i$ its normalized time.  Its residual relative
to the endpoint chord is defined in \eqref{eq:trajectory-residual}:

\begin{equation}
  \label{eq:trajectory-residual}
  r_i=p_i-\left[p_0+u_i(p_N-p_0)\right].
\end{equation}

Here, $p_0+u_i(p_N-p_0)$ is the point at the same normalized time on the
straight chord from the reference start to end, so $r_i$ captures the local
deviation of the human trajectory from that chord.

Given the start point $s$ and endpoint $e$ requested by the agent, ActReal
generates the transformed point in \eqref{eq:trajectory-transform}:

\begin{equation}
  \label{eq:trajectory-transform}
  \hat p_i=s+u_i(e-s)+\alpha A r_i,
\end{equation}

where $A$ rotates and scales the reference's local coordinate frame to the
requested direction and distance.  The global factor $0\leq\alpha\leq1$ keeps
every point on screen.  This transform exactly realizes the requested endpoints while
retaining the reference trajectory's curvature, local jitter, and velocity
variation.

The two actions differ primarily in timing.  ActReal determines a scroll's
duration by interpolating, in log space, the travel--duration pairs in its five
references, and uses the same duration to select the corresponding IMU interval.
For a swipe, ActReal instead recovers a base report period and skip pattern from
the reference, then synthesizes a new irregular update cadence.  When the
requested event requires additional updates, it interpolates reference
coordinates and pressure before mapping the updates to a common observation
clock, retaining the rhythm of the stroke and release phase.

\paragraph{Pinch.}
A pinch is constrained by its target region and inter-finger span. Its opening
or closing direction is inherited from the selected reference.  ActReal first
selects a five-shot reference whose maximum span is close to that requested.
Let $c^\ast$ be the reference center at its maximum-span frame and $c_T$ the
center of the target region.  At the design level, the raw
two-pointer mapping is the similarity transform in \eqref{eq:pinch-transform}:

\begin{equation}
  \label{eq:pinch-transform}
  \hat q_{i,k}=c_T+\beta R\left(q_{i,k}-c^\ast\right),
\end{equation}

where $q_{i,k}$ is the position of finger $k$ at time $i$, $R$ rotates the
reference finger axis onto the target-region axis, and $\beta$ is the ratio of
the requested span to the reference maximum span.  Applying the same $R$ and
$\beta$ throughout the gesture preserves its relative span changes and center
motion.
The current offline evaluation retains only the two-finger centroid $c_i$
visible to the detector.  Its output therefore uses the equivalent transform
in \eqref{eq:pinch-centroid-transform}:

\begin{equation}
  \label{eq:pinch-centroid-transform}
  \hat c_i=c_T+\beta R(c_i-c^\ast).
\end{equation}

\paragraph{Keystroke.}
Given a \texttt{type(text)} request, ActReal uses the current keyboard layout to map the requested text to a sequence of key coordinates $a_1,\ldots,a_m$. ActReal collects key-hold durations $h_j$ and inter-key
pauses $f_j$ from the five keystroke touch references, cycles through these timings in their recorded order, and projects them onto the requested event duration.

For key $j$, let $e_{j,\ell}$ be a reference press's local motion relative to
its landing point.  ActReal generates its position using
\eqref{eq:keystroke-position}:

\begin{equation}
  \label{eq:keystroke-position}
  p_{j,\ell}=a_j+\gamma e_{j,\ell},
\end{equation}

where $0\leq\gamma\leq1$ limits the within-key excursion to Android's
touch-slop range so that a press is not interpreted as a drag. The motion sequence expressed in
\eqref{eq:keystroke-schedule} accumulates the key-hold durations and inter-key pauses:

\begin{equation}
  \label{eq:keystroke-schedule}
  \tau_1=0,
  \qquad
  \tau_{j+1}=\tau_j+h_j+f_j.
\end{equation}

The touch composer passes this sequence to the keystroke IMU adapter, placing
each key contact and its associated device motion at the same time.

\subsection{System Event Delivery}
\label{sec:method-injection}
ActReal uses a unified system-level execution interface to translate semantic
actions generated by the agent into an \texttt{ActionBundle} containing touch
and IMU signals. Actions generated by different mobile-agent frameworks are
first standardized into action types supported by ActReal and are then executed
through the same underlying mechanism.

For touch input, the ActReal prototype registers a persistent multi-touch
\texttt{uinput} device. The device configures the relevant touch axes and their
value ranges based on the physical touchscreen and is declared as a direct
touchscreen, allowing the generated events to traverse the standard Android
input stack and reach the foreground application as \texttt{MotionEvent}s.

For the IMU, ActReal performs runtime native spoofing within the target
application process. After a batch of physical sensor events is returned from
\texttt{SensorEventQueue::read}, the hook replaces only the three-axis values of
the accelerometer or gyroscope while preserving the original sensor handle,
sensor type, timestamp, sampling configuration, and batching behavior.
Consequently, the application still sees the physical sensor objects and their
metadata through \texttt{SensorManager}. This mechanism does not modify the
target APK on disk, nor does it require the application to integrate the
ActReal SDK.

Because touch and IMU are delivered through different Android event paths, ActReal aligns their clocks before execution and maps the unified \texttt{ActionBundle} timeline onto both streams, ensuring temporal consistency
between the touch and IMU signals associated with the same action. Agent-side operations such as screen capture, OCR, and model inference can introduce long
idle intervals. During these periods, ActReal continuously provides background IMU signals and inserts short filler taps to maintain continuous human-like
interaction behavior. These filler taps are placed in non-interactive regions that do not trigger UI actions and are scheduled with corresponding IMU signals
independently of the agent's task actions. In this way, ActReal preserves the agent's original planning and inference delays while avoiding long gaps in
observable touch or sensor activity and maintaining a common time base between touch and IMU throughout the session.

\section{Evaluation}
\label{sec:evaluation}

We evaluate ActReal along three dimensions: attack effectiveness, real-device execution, and design contributions. We first report event- and session-level attack results and baseline comparisons, then evaluate ActReal over real Android event paths and measure injection latency, and finally use ablation studies to analyze the contributions of five-shot target-user conditioning, adversarial training, and structural constraints.

\subsection{Experimental Setup}
\label{sec:eval-setup}

\paragraph{Dataset and User Split.}
For evaluation of our system, we adopt the open-source dataset HMOG, which contains synchronized touch and device motion data in human-device interactions~\cite{sitova2015hmog}.  HMOG was collected with ten Samsung Galaxy S4
phones from 100 participants performing reading, text-entry, and map-navigation
tasks~\cite{yang2014multimodal}.  It
records timestamped raw Android touch events and samples the three-axis
accelerometer, gyroscope, and magnetometer at 100~Hz. We use the touch,
accelerometer, and gyroscope signals, which are necessary for our system construction.

We partition the 100 HMOG users into a fixed, mutually disjoint split of 70 training users, 10 validation users, and 20 test users, shared by all actions. The training users are used to learn the generators and behavioral detectors, the validation users to determine detector thresholds, and the test users to report attack results. For each target user and action, ActReal freezes five genuine reference interactions before generation. It generates 200 events for each user--action pair, giving 14,000 training events, 2,000 validation events, and 4,000 test events per action. All methods use the same user split, genuine events, and preprocessing.

\paragraph{Defender observations and models.}
We consider three app-visible observation modalities.  The touch detector reads
a nine-channel trajectory containing contact state, XY coordinates, pressure,
pointer count, coordinate changes, event time, and field availability.  The IMU
detector reads only three-axis acceleration and three-axis angular velocity.
The joint detector reads all 15 channels on the same timeline and can therefore
observe touch coordinates, touch timing, and the corresponding device motion.
However, for Keystroke input, touch events on the system keyboard are received
by the input method editor (IME) and converted into text-edit events. Therefore,
the target application can observe only the editing results and cannot observe
the raw touch events on the keyboard. The Keystroke Touch and Joint ASRs reported
in this paper serve only as diagnostic references for ActReal-generated signals.

The detector suite contains six models: HMOG-SVM, HMOG-RF, TToS-SVM,
TToS-XGB, BehaveFormer, and AuthConFormer. HMOG-SVM and HMOG-RF are
HMOG-inspired~\cite{sitova2015hmog} statistical models that use a linear SVM and
a random forest. TToS-SVM and TToS-XGB adopt the linear-SVM and XGBoost classification
frameworks evaluated by the mobile-agent humanization benchmark \emph{Turing
Test on Screen}~\cite{zhu2026turing}; their inputs combine
Touchalytics-style~\cite{frank2012touchalytics} touch-dynamics features with
RiskCog-derived~\cite{zhu2019riskcog} inertial statistics. BehaveFormer and
AuthConFormer are adapted from mobile-user authentication: BehaveFormer uses a
temporal--channel dual-attention architecture~\cite{senarath2023behaveformer},
whereas AuthConFormer uses a convolution--Transformer
architecture~\cite{hu2023authconformer}; both
operate directly on raw event sequences.  All six models are retrained as
global human-versus-generated binary classifiers.

We use an attack-aware adaptive-defense setting.  For ActReal and each
comparison method, the detector for each action is trained independently using
genuine training events from the 70 training users and 14,000 generated training
events from that method. The validation set contains genuine validation events
from the 10 validation users and 2,000 generated validation events for threshold
selection; after freezing, testing uses genuine test events from the 20 test
users and 4,000 generated test events.

\paragraph{Thresholds and metrics.}
We use \emph{attack success rate} (ASR) as our primary metric. An attack
succeeds when the detector accepts an ActReal-generated interaction as human.
Event-level ASR is the fraction of generated events accepted as human, whereas
session-level ASR is the fraction of attack sessions accepted after applying
the specified score-aggregation rule. For both evaluation granularities, each
detector is trained only on the training users. Following common industry
practice, we select the operating threshold on the validation users under a
maximum false rejection rate (FRR) of 5\% on genuine validation
events~\cite{googleTransactionDefense}. Among the thresholds satisfying this constraint, we choose the one that minimizes validation-set ASR; ties are
broken by selecting the threshold whose FRR is closest to 5\%. The detector, preprocessing, aggregation rule, and threshold are then frozen before evaluation on the unseen test users.

\subsection{Main Attack Effectiveness}
\label{sec:eval-main}

\paragraph{Overall Attack Effectiveness.}
To evaluate ActReal's effectiveness against behavioral presentation attack detection (PAD), we measure event-level ASR across five actions, three observation modalities, and six detectors, covering 90 action--modality--detector combinations. Across the full evaluation grid, ActReal achieves a macro-averaged ASR of 77.5\% (user-clustered 95\% CI: [76.3\%, 78.7\%]). The corresponding FRR on genuine
test events is 5.2\% (95\% CI: [4.6\%, 5.9\%]), confirming that the predefined operating point remains stable on unseen users.
Table~\ref{tab:actreal-breakdown} reports the event-level ASR for all action--modality--detector combinations, while Table~\ref{tab:actreal-aggregate} summarizes aggregate ASR and FRR with user-clustered confidence intervals.

\begin{table*}[t]
  \centering
  \small
  \setlength{\tabcolsep}{2.2pt}
  \renewcommand{\arraystretch}{1.12}
  \caption{ActReal event-level ASR (\%) for all 90 action--modality--detector
  cells. T, I, and J denote touch trajectory, six-axis IMU, and joint
  observation. The Mean ASR row macro-averages the five actions.}
  \label{tab:actreal-breakdown}
  \begin{tabular}{@{}l*{6}{ccc}@{}}
    \toprule
    & \multicolumn{3}{c}{HMOG-SVM~\cite{sitova2015hmog}}
    & \multicolumn{3}{c}{HMOG-RF~\cite{sitova2015hmog}}
    & \multicolumn{3}{c}{TToS-SVM~\cite{zhu2026turing}}
    & \multicolumn{3}{c}{TToS-XGB~\cite{zhu2026turing}}
    & \multicolumn{3}{c}{BehaveFormer~\cite{senarath2023behaveformer}}
    & \multicolumn{3}{c}{AuthConFormer~\cite{hu2023authconformer}} \\
    \cmidrule(lr){2-4}\cmidrule(lr){5-7}\cmidrule(lr){8-10}
    \cmidrule(lr){11-13}\cmidrule(lr){14-16}\cmidrule(l){17-19}
    Action & T & I & J & T & I & J & T & I & J
           & T & I & J & T & I & J & T & I & J \\
    \midrule
    Tap       & 90.7 & 88.4 & 88.6 & 86.6 & 92.2 & 83.5
              & 89.6 & 90.2 & 88.3 & 91.5 & 85.4 & 79.7
              & 91.6 & 90.4 & 74.4 & 90.3 & 82.1 & 77.4 \\
    Scroll    & 75.9 & 82.8 & 70.9 & 61.5 & 70.7 & 48.4
              & 76.5 & 80.6 & 71.5 & 64.2 & 68.0 & 48.2
              & 75.4 & 70.7 & 72.8 & 57.3 & 62.8 & 52.2 \\
    Swipe     & 74.2 & 89.0 & 74.7 & 62.7 & 83.7 & 55.7
              & 79.0 & 89.8 & 74.1 & 70.7 & 77.0 & 59.8
              & 76.9 & 84.4 & 67.6 & 50.2 & 87.1 & 59.8 \\
    Pinch     & 70.5 & 90.9 & 69.6 & 54.4 & 82.1 & 54.2
              & 79.0 & 91.8 & 80.6 & 55.9 & 77.2 & 52.3
              & 84.1 & 93.6 & 86.1 & 93.5 & 94.3 & 89.7 \\
    Keystroke & 94.4 & 71.2 & 73.9 & 94.9 & 81.9 & 82.2
              & 83.7 & 81.2 & 79.6 & 74.2 & 79.2 & 66.3
              & 94.6 & 95.0 & 96.1 & 88.3 & 91.4 & 55.6 \\
    \midrule
    \rowcolor{MeanBlue}
    \textbf{Mean} & \textbf{81.2} & \textbf{84.5} & \textbf{75.5}
              & \textbf{72.0} & \textbf{82.1} & \textbf{64.8}
              & \textbf{81.6} & \textbf{86.7} & \textbf{78.8}
              & \textbf{71.3} & \textbf{77.4} & \textbf{61.3}
              & \textbf{84.5} & \textbf{86.8} & \textbf{79.4}
              & \textbf{75.9} & \textbf{83.5} & \textbf{67.0} \\
    \bottomrule
  \end{tabular}

\end{table*}

\begin{table}[t]
  \centering
  \small
  \caption{Event-level ASR (\%) on generated test events and FRR (\%) on genuine test events from unseen test users.}
  \label{tab:actreal-aggregate}
  \begin{tabular}{@{}lcc@{}}
    \toprule
    Metric & Estimate & 95\% CI \\
    \midrule
    Event-level ASR    & 77.5  & [76.3, 78.7] \\
    FRR                & 5.2   & [4.6, 5.9] \\
    \bottomrule
  \end{tabular}
\end{table}

The results reveal two consistent trends. First, attack success depends on the
observed modality: averaged across actions and detectors, ASR is highest for
IMU-only observation (83.5\%), followed by touch-only (77.7\%), and lowest
under joint touch--IMU observation (71.1\%). Thus, combining touch and IMU
provides a stronger detection signal, although ActReal still achieves an
average ASR above 70\% under joint observation. Second, effectiveness varies
across actions. Tap achieves the highest action-level ASR (86.7\%), whereas
Scroll is the most challenging (67.3\%); the lowest individual results occur
for joint Scroll under TToS-XGB (48.2\%) and HMOG-RF (48.4\%).

\paragraph{Session-Level Detection.}
In practical deployments, some applications continuously accumulate behavioral
evidence over an interaction session in addition to judging individual
interaction events, and use it to form an overall risk alert. However, the
occurrence times and action densities of mobile interactions are not fixed. When
a fixed-length time window is used to train a temporal detector, a short window
may fail to contain enough scoreable events, while a complete action that crosses
a window boundary and its corresponding IMU response may be split; a long window,
in contrast, increases detection latency and makes it difficult to use a uniform
length across tasks such as reading, text entry, and navigation. We therefore
adopt session-level score aggregation over complete interaction events, without
assuming a fixed event-arrival rate or splitting an action's touch--IMU evidence
at fixed time-window boundaries.

We construct attack sessions from 474 genuine sessions of 20 test users,
preserving each session's target user, number of events, and action composition.
The frozen detectors assign a generated-class score to each event, which we aggregate using
three rules:
\emph{Count} counts events exceeding the action-specific 5\% FRR threshold,
\emph{Mean} averages all event scores, and \emph{Trimmed Mean} averages the
highest-scoring one-third of events to emphasize the most suspicious events. To
reduce sensitivity to random sampling, we repeat session construction 20 times
and evaluate 200 balanced 10/10 user partitions for threshold selection and
testing.

As shown in Table~\ref{tab:session-bypass}, ActReal achieves session-level ASRs
of 63.5\%, 51.2\%, and 50.5\% under Count, Mean, and Trimmed Mean. Session-level aggregation lowers attack success compared with
event-level evaluation, but more than half of the attack sessions remain
accepted under all three aggregation rules.

\begin{table}[t]
  \centering
  \small
  \caption{Session-level ASR (\%) and FRR (\%) under three score-aggregation rules; FRR is computed over genuine sessions.}
  \label{tab:session-bypass}
  \begin{tabular}{@{}lccc@{}}
    \toprule
    Aggregation & FRR & Session ASR & 95\% CI \\
    \midrule
    Count        & 4.0 & \textbf{63.5} & [58.2, 67.2] \\
    Mean         & 5.2 & 51.2          & [43.7, 57.3] \\
    Trimmed mean & 5.2 & 50.5          & [43.9, 57.1] \\
    \bottomrule
  \end{tabular}
\end{table}

\paragraph{Comparison with Existing Generators}
We next compare ActReal with representative time-series and trajectory generators. Among the existing methods that we could obtain and reproduce, we found no method that uses five target-user references through the same interface while generating both touch and IMU.  We therefore select
Diffusion-TS~\cite{yuan2024diffusion} and
ImagenTime~\cite{naiman2024utilizing} to cover two diffusion-based generation
routes: the former directly models and decomposes a time series, whereas the latter converts it into an image through delay embedding before diffusion.
TTS-GAN~\cite{li2022tts} represents Transformer-based adversarial generation. For touch trajectories, we select the B\'{e}zier-curve-based pyclick~\cite{pyclick} and ghost-cursor~\cite{ghostcursor}, which combines Fitts'law, overshoot, and correction segments, to cover training-free trajectory
humanization methods that can be placed directly in an executor. And Diffusion-TS also generates touch, IMU, and both streams together, providing a common
comparison across all three observation modalities.

\begin{figure}[t]
  \centering
  \includegraphics[width=\columnwidth]{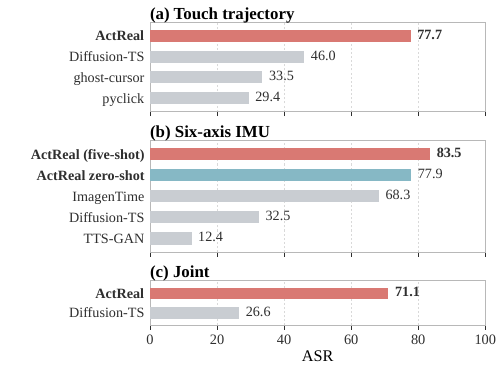}
  \caption{Event-level ASR of ActReal and existing generators under
(a) touch, (b) IMU, and (c) joint observation. Results are macro-averaged
over actions and detectors.}
  \label{fig:generator-comparison}
\end{figure}

Figure~\ref{fig:generator-comparison} shows that ActReal consistently outperforms existing generators across all three observation modalities, achieving ASRs of
77.7\%, 83.5\%, and 71.1\% for touch, IMU, and joint observation. The strongest baselines reach only 46.0\%, 68.3\%, and 26.6\% and even removing five-shot target-user conditioning reduces ActReal's IMU ASR from 83.5\% to
77.9\%, confirming the benefit of target-user references.

We find that the weaker baseline performance mainly comes from detectable waveform artifacts. Diffusion-TS has high lag-1 autocorrelation (0.951--0.996) but only 0.561--0.762 of the genuine standard deviation, indicating overly smooth signals. TTS-GAN better matches signal magnitude (standard-deviation ratio: 0.949--1.084) but shows much lower lag-1
autocorrelation for Scroll, Swipe, and Keystroke (0.729, 0.745, and 0.364, versus 0.996, 0.996, and 0.971 for genuine events), indicating weaker temporal continuity. These results support ActReal's design, which reduces the waveform artifacts that limit existing generators.

\subsection{Evaluation on Real Devices}
\label{sec:eval-device}
\paragraph{Attack Effectiveness.}
To evaluate ActReal over real Android event paths, we conduct experiments on a
Galaxy S21 and a Pixel~10 with 20 new
participants. Each participant manually completed three controlled
tasks---Shopping, Search/Reading, and Social---on both phones.
Before performing the tasks, each participant provided five synchronized
touch--IMU references for each of Tap, Scroll, Swipe, Pinch, and Keystroke.
These references are used only to condition signal generation and are excluded
from test events.

\begin{figure}[t]
  \centering
  \begin{minipage}[t]{0.48\columnwidth}
    \centering
    \includegraphics[width=\linewidth,trim=0 534 0 100,clip]
      {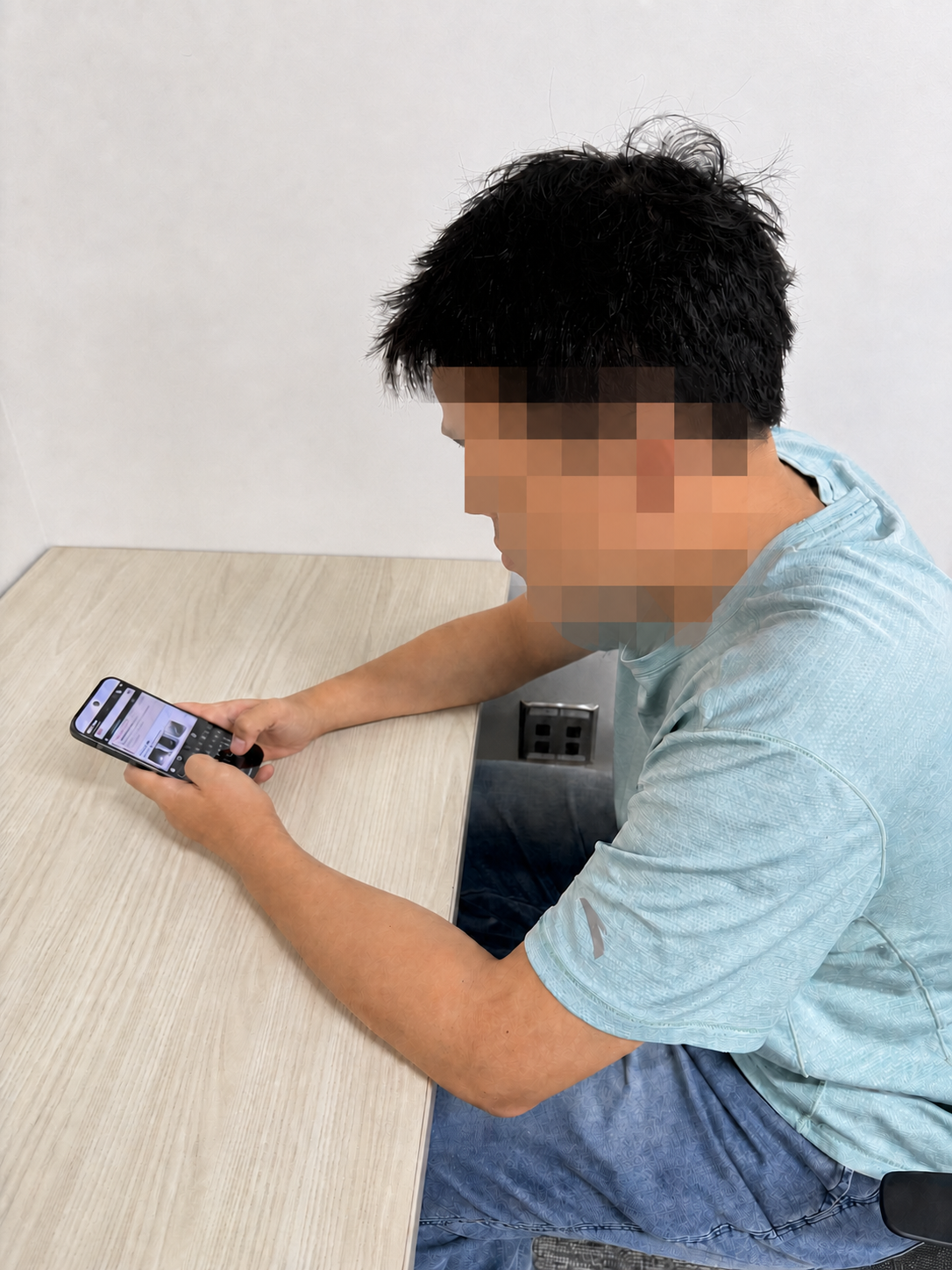}
  \end{minipage}\hfill
  \begin{minipage}[t]{0.48\columnwidth}
    \centering
    \includegraphics[width=\linewidth]
      {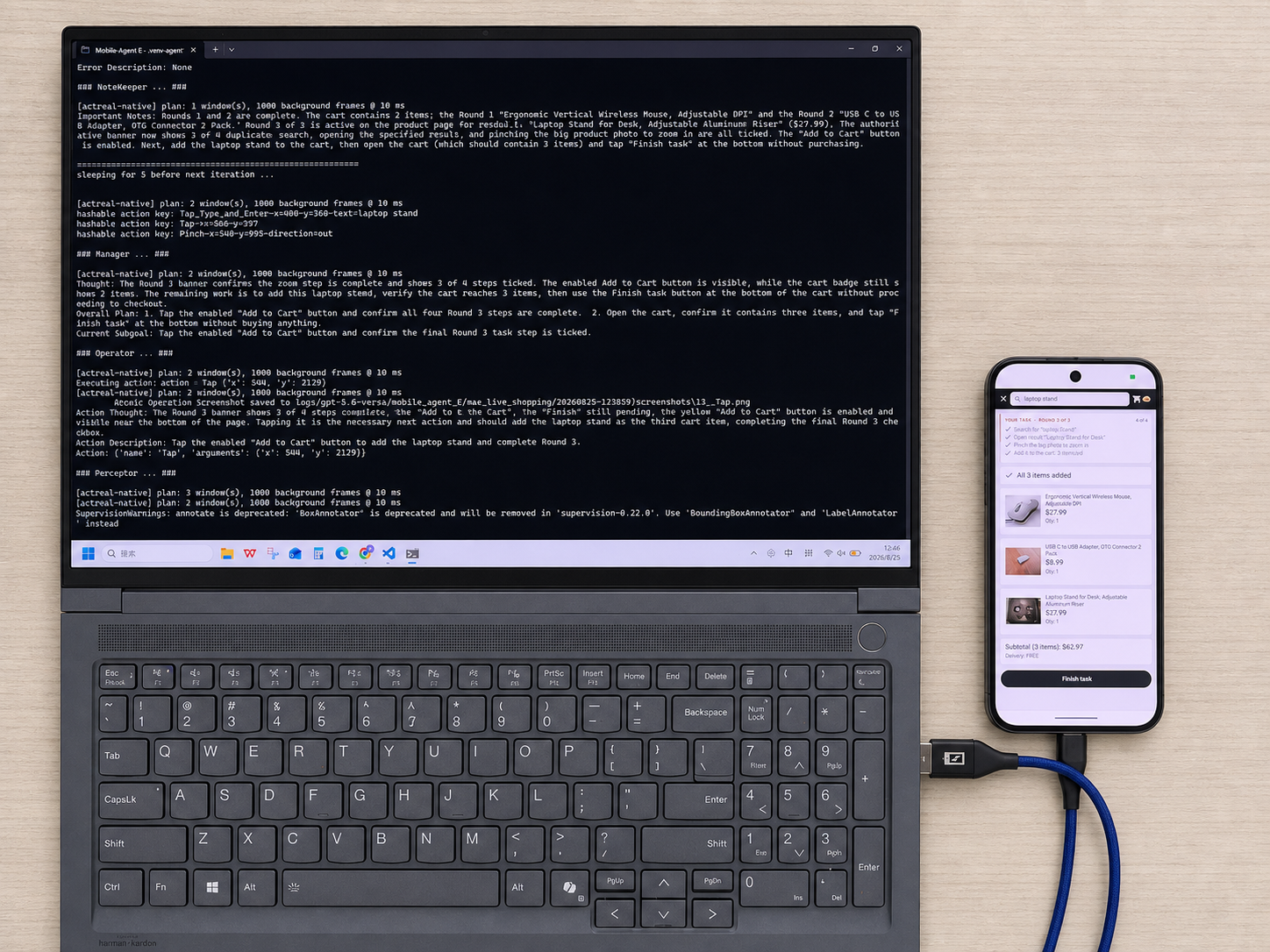}
  \end{minipage}
  \caption{Real-device evaluation setup. Left: a participant completes a
  controlled task on a test phone (face pixelated for privacy). Right: a
  mobile agent executes the same task through the system-level executor.}
  \label{fig:real-device-setup}
\end{figure}

The mobile-agent evaluation includes Mobile-Agent-E~\cite{wang2025mobile}, AppAgent~\cite{zhang2025appagent}, and MobileUse~\cite{li2026mobileuse}, all using GPT-5.6 Terra~\cite{openaiGPT56Terra} as the action-decision model. Mobile-Agent-E is limited to 25 steps on the Galaxy S21 and, on the Pixel 10, to 25 steps for Shopping and 40 steps for Search/Reading and Social; AppAgent and MobileUse are limited to 60 and 90 steps, respectively. We reuse the frozen detector models together with their preprocessing pipelines and decision thresholds. The evaluation covers five actions, three observation modalities, and six detectors. During execution, we score all application-visible events, including filler taps. For each action and modality, ASR is computed by aggregating the decisions of all six detectors over the corresponding scoreable events. For genuine keystrokes, as noted above, our simulated application cannot observe the corresponding touch events on the system keyboard during data collection. Therefore, we do not report the corresponding Touch or Joint FRR values. Figure~\ref{fig:real-device-setup} illustrates the participant-side and agent-side real-device evaluation setups.

\begin{table}[t]
  \centering
  \caption{Real-device ASR (\%) and FRR (\%), aggregated over six detectors.}
  \label{tab:device-results}
  \small
  \begin{tabular*}{\columnwidth}
    {@{\extracolsep{\fill}}lrrrrrr@{}}
    \toprule
           & \multicolumn{2}{c}{Touch}
           & \multicolumn{2}{c}{IMU}
           & \multicolumn{2}{c}{Joint} \\
    \cmidrule(lr){2-3}
    \cmidrule(lr){4-5}
    \cmidrule(l){6-7}
    Action & ASR & FRR & ASR & FRR & ASR & FRR \\
    \midrule
    Tap       & 95.3 &  8.0 & 91.5 &  9.1 & 88.3 & 17.6 \\
    Scroll    & 71.5 &  9.6 & 86.8 &  3.6 & 69.2 &  5.0 \\
    Swipe     & 53.7 &  3.5 & 94.4 & 14.5 & 48.1 &  5.2 \\
    Pinch     & 88.3 &  2.9 & 79.0 & 17.2 & 78.8 & 14.3 \\
    Keystroke & 77.8 & \multicolumn{1}{c}{--}
              & 81.3 &  6.0
              & 61.8 & \multicolumn{1}{c}{--} \\
    \bottomrule
  \end{tabular*}
\end{table}

Table~\ref{tab:device-results} reports ASR and FRR. Although joint observation generally reduces ASR, it still accepts 48.1--88.3\% of ActReal events as genuine, showing that ActReal remains effective against the frozen detectors on real devices. Meanwhile, FRR on genuine events varies substantially across actions and modalities, reaching up to 17.6\%. In particular, joint observation does not consistently reduce FRR and can reject more genuine events for some actions, highlighting the practical tradeoff between stricter multimodal detection and usability.

\paragraph{Execution overhead.}
We first examine whether enabling ActReal affects successful task execution. Across all 18 agent--task--device configurations, the agents complete the assigned tasks successfully with ActReal enabled, suggesting that the additional injection layer does not disrupt normal execution. We further quantify the runtime overhead introduced by ActReal's touch-injection path as the time between releasing an event through the persistent multi-touch \texttt{uinput} device and observing the first corresponding \texttt{MotionEvent} in the test application. Across 20 probes per device, the median latency is 10.35\,ms on the Pixel~10 and 8.99\,ms on the Galaxy S21, with P95 latencies of 13.93\,ms and 11.43\,ms, respectively. As shown in Table~\ref{tab:device-latency}, ActReal's first-touch-event delivery latency is small and stable at the millisecond scale and is negligible relative to the perception and reasoning latency of mobile agents.
\begin{table}[t]
  \centering
  \caption{First-touch-event delivery latency on real devices. Each result is measured
  over 20 probes.}
  \label{tab:device-latency}
  \small
  \begin{tabular*}{.6\columnwidth}
    {@{\extracolsep{\fill}}lrr@{}}
    \toprule
    Device & Median (ms) & P95 (ms) \\
    \midrule
    Pixel~10   & 10.35 & 13.93 \\
    Galaxy S21 &  8.99 & 11.43 \\
    \bottomrule
  \end{tabular*}
\end{table}

\begin{figure}[t]
  \centering
  \includegraphics[width=\columnwidth]{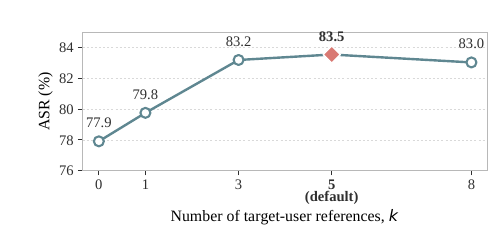}
  \caption{Reference-count ablation for ActReal's IMU generator. ASR is macro-averaged over four gesture actions and six IMU-only detectors; $k=5$ is the default setting in ActReal.}
  \label{fig:ablation-reference}
\end{figure}

\subsection{Ablation Study}
\label{sec:eval-ablation}

To analyze the contributions of target-user references and IMU training components to attack effectiveness, we ablate ActReal's IMU branch. For each
setting, we regenerate the attack events, train the corresponding detectors, and reselect thresholds satisfying the 5\% FRR requirement on the
validation users. We separately vary the reference count and remove the three critics, feature matching, and the structural constraints and protected
updates. Keystroke is excluded because it uses a separate non-diffusion IMU generator.

Figure~\ref{fig:ablation-reference} compares attack effectiveness under different numbers of reference samples. All results are averaged over 24 configurations,
comprising four action classes---tap, scroll, swipe, and pinch---and six IMU-only detectors. ActReal uses five target-user references by default; the measured ASR rises from 77.9\% without references to 83.5\% at $k=5$,
while the observed results at $k=3$, $k=5$, and $k=8$ remain within 0.6 percentage points.

Table~\ref{tab:ablation-components} evaluates the contribution of each training component by removing the feature, set, and waveform critics individually. We also evaluate a variant without the structural losses and protected-update mechanisms. All results are averaged over the same evaluation and all rows are averaged over the same 24 configurations comprising tap, scroll, swipe, and pinch with six IMU-only detectors.

\begin{table}[t]
  \centering
  \small
  \caption{Ablation of ActReal's IMU training components. Results are averaged over four gesture actions and six IMU-only detectors; $\Delta$ denotes the ASR change in percentage points.}
  \label{tab:ablation-components}
  \begin{tabular}{@{}lrrr@{}}
    \toprule
    Removed component & Base & Ablated & $\Delta$ (pp) \\
    \midrule
    Feature critic                 & 83.5 & 71.4 & $-12.1$ \\
    Set critic                     & 83.5 & 74.1 & $-9.5$ \\
    Waveform critic                & 83.5 & 73.4 & $-10.2$ \\
    Structure + protected updates  & 83.5 & 56.1 & $-27.4$ \\
    \bottomrule
  \end{tabular}
\end{table}

\section{Related Work}
\label{sec:related-work}

\subsection{Mobile Automation}

In recent years, large language models (LLMs) and multimodal large language
models have increasingly been used to build mobile GUI agents that can
autonomously complete complex, multi-step tasks based on user instructions and
the live interface state. A typical agent operates in a
perception--reasoning--execution--feedback loop. It first observes the current
state through screenshots or UI elements, generates the next action based on
the user's goal and its action history, and then continues reasoning over the
resulting interface~\cite{yao2022react}.
On mobile devices, AutoDroid uses the current UI and app-specific knowledge to
guide LLM action generation~\cite{wen2024autodroid}; AppAgent acquires knowledge
about how to operate applications through autonomous
exploration~\cite{zhang2025appagent}; and Mobile-Agent relies primarily on visual
perception to identify interface elements and plan actions
incrementally~\cite{wang2024mobile}.

By contrast, conventional mobile automation typically relies on tools such as
Appium, UIAutomator, and the Android Debug Bridge (ADB) to execute prewritten
scripts. Developers must specify the target UI elements, actions, and execution
order in advance. Changes to the UI layout or widget state can invalidate
existing scripts and require manual maintenance~\cite{coppola2019fragility}.
Conventional scripts therefore execute actions according to predefined rules,
whereas mobile GUI agents select each next action online based on the user's
goal, the current interface, and prior interaction outcomes. The main
distinction lies at the decision layer.

\subsection{System-Level Agents and Execution Privileges}

An agent's deployment and system privileges determine how model-generated
actions are executed on a device. Existing mobile agents generally use one of
two execution models. The first uses external automation channels, such as ADB
or the Android Accessibility Service, to observe the interface and execute
actions. Research prototypes including AutoDroid, AppAgent, and Mobile-Agent
primarily adopt this model~\cite{wu2025assistants}. The second integrates the
agent as a preinstalled or privileged system component, allowing it to invoke
private system APIs and use platform-signature permissions to capture screen
content, launch applications, or inject input events directly~\cite{zou2026blind}.
For example, AOHP builds an OS-level agent harness on top of AOSP, treats the
agent as a first-class actor in the operating system, and provides native
execution interfaces and secure information-flow mechanisms~\cite{zhao2026aohp}.

Android's conventional security model isolates applications and their data
primarily through application sandboxing and permissions~\cite{felt2011android}.
To complete cross-application tasks, however, a system-level agent requires
capabilities for screen perception, application launching, cross-application
information access, and action execution. These capabilities broaden the
security boundary that the conventional permission model must protect. Using
the Doubao Mobile Assistant as a representative system-integrated deployment,
Zou et al. analyze security risks involving agent identity, external interfaces,
internal reasoning, and action execution. They propose Aura, which enforces
fine-grained permission control through a System Agent, sandboxed App Agents,
and an Agent Kernel~\cite{zou2026blind}. Wu et al. further analyze nine
mobile agents and identify eleven classes of attack surfaces spanning reasoning,
GUI interaction, and system execution~\cite{wu2025assistants}. \emph{Mind the
Gap} demonstrates how a malicious application can exploit the gap between an
agent's observation and execution to rebind a planned action to a sensitive
control in another application~\cite{qian2026mind}. These studies primarily
examine system privileges and action authorization. In contrast, this paper
asks whether, after an action reaches the target application, the application
can distinguish an agent from a human solely from the interaction behavior it
observes.

\subsection{Application-Side Automation Detection}

In the application-side automation-detection setting considered in this paper,
the target application analyzes interaction events. From these
events, the application can derive two broad classes of features. Action-level
features include the action type---such as a tap, swipe, or text input---its
coordinates, the action sequence, inter-action intervals, and press duration.
Trajectory-level features capture the timestamped sequence of coordinates
formed by touch-down, touch-move, and touch-up events, together with derived
measures such as path length, straightness, curvature, velocity, acceleration,
and jerk.

Prior work on touch-based behavioral authentication has shown that touch
trajectories observed by an application can carry discriminative behavioral
information. Touchalytics extracts 30 behavioral features from raw touch logs to
distinguish users based on their swipe patterns~\cite{frank2012touchalytics}.
BeCAPTCHA further uses the touch trajectory and accelerometer signals associated
with a single drag-and-drop gesture to distinguish humans from automated bots.
It also constructs synthetic bot samples using both handcrafted methods and
generative adversarial networks (GANs)~\cite{acien2021becaptcha}. Together, these
studies indicate that touch trajectories and their kinematic properties can
provide useful signals for identifying non-human interaction.

\section{Discussion}
\label{sec:discussion}

\paragraph{Limitations.}
ActReal primarily targets ordinary applications that rely on touch and IMU
signals for passive behavioral detection; it does not target high-assurance
applications such as banking, payment, and identity-authentication services.
These high-assurance applications typically employ multiple layers of security
mechanisms, including biometric authentication, application and device
integrity checks, protected transaction confirmation, and backend risk
controls. These mechanisms do not directly depend on the authenticity of touch
and IMU signals. Therefore, even if ActReal generates highly realistic touch
and IMU signals, its operations may still be identified, blocked, or subjected
to additional verification by other security mechanisms.

\paragraph{Defensive Implications.}
Future defenses can be developed at both the agent and application levels. At
the agent level, the system should follow the principle of least privilege and
restrict an agent executor's access to the input-injection and sensor-delivery
paths, particularly preventing the same executor from controlling both
capabilities. For high-risk privileges, the system can additionally require
separate authorization, explicit user confirmation, operation auditing, and
timely revocation. At the application level, applications should not rely
solely on local touch and IMU signals. Instead, they should incorporate
server-side information and construct a comprehensive risk score based on
access frequency, network environment, device--account associations,
geographic-location changes, business-operation patterns, and historical risk
records.
\section{Conclusion}
\label{sec:conclusion}
Mobile agents are increasingly moving beyond external automation mechanisms, such as ADB and the Accessibility Service, toward deeper integration with system-level components of mobile devices. While this transition enables greater autonomy and more flexible task execution, it also fundamentally changes the security boundary. In particular, when a privileged executor can control both Android's input-dispatch and sensor-delivery paths, applications can no longer rely on touch--IMU consistency as trustworthy evidence of human interaction. To investigate this emerging risk, we develop ActReal, which maps heterogeneous agent operations into five physical action classes and realizes human-like interactions through two distinct execution paths: touch events are injected through Android's input stack, while application-visible IMU values are replaced through a native hook; the two streams remain aligned on a shared timeline. Across three observation modalities and six detectors, ActReal achieves a mean event-level attack success rate of 77.5\%, while maintaining an ASR of 71.1\% under joint touch--IMU observation. At the session level and across different real devices, ActReal remains effective under stricter detection settings, demonstrating that the attack persists through actual Android execution paths. These results highlight a fundamental challenge that system-level mobile agents pose to existing mobile automation detection.

\bibliographystyle{plainurl}
\bibliography{references}

\end{document}